\documentclass[useAMS,usenatbib]{mn2e}

\usepackage{graphicx}

\newcommand{\Tef}{T$_{\rm eff}$~}

\newcommand{\Vt}{$V_t$~}

\newcommand{\vsini}{$v$ sin $i$}

\newcommand{\grad}{$^o$~}

\title[Helium and silicon in HD 21699]{Spots structure and stratification of helium and silicon \\ in the atmosphere of He-weak star HD 21699.}
\author[A. Shavrina et al.]
{A.V.Shavrina$^{1}$, Yu.V. Glagolevskij$^{2}$, J. Silvester$^{3,4}$,
\newauthor {G.A. Chuntonov$^{2}$, V.R. Khalack$^{5}$, Ya.V. Pavlenko$^{1}$}\\
$^{1}$Main Astronomical Observatory of National Academy of Sciences of Ukraine,
 27 Akademika Zabolotnoho St., 03680 Kyiv, Ukraine \\
$^{2}$Special Astrophysical Observatory of Russian Academy of Sciences,
Nizhnij Arkhyz, Zelenchukskiy region, Karachai-Cherkessian Republic, 
 Russia 369167 \\
$^{3}$Department of Physics, Engineering Physics \& Astronomy, Queen's University, Kingston, Ontario, Canada, K7L 3N6\\
$^{4}$Department of Physics, Royal Military College of Canada, P.O. Box 17000, Station `Forces', Kingston, Ontario, Canada, K7K 7B4\\
$^{5}$D\'{e}partement de Physique et d'Astronomie,Universit\'{e} de Moncton, Moncton, N.-B., Canada E1A 3E9  }

\begin{document}

\date{}
\pagerange{\pageref{firstpage}--\pageref{lastpage}} \pubyear{2002}

\maketitle

\label{firstpage}

\begin{abstract}

The magnetic star HD 21699 possesses a unique magnetic field structure where
the magnetic dipole is displaced from the centre by 0.4 $\pm$ 0.1 of
the stellar radius (perpendicularly to the magnetic axis), as a
result, the magnetic poles are situated close to one another on the
stellar surface with an angular separation of 55\grad and not
180\grad  as seen in the case of a centred dipole. Respectively, the
two magnetic poles form a large "magnetic spot". High-resolution
spectra were obtained allowing He I and Si II abundance variations to
be studied as a function of rotational phase. The results
show that the helium abundance is concentrated in one hemisphere of
the star, near the magnetic poles and it is comparatively weaker in
another hemisphere, where magnetic field lines are horizontal with
respect to the stellar surface. At the same time, the silicon
abundance is greatest between longitudes of 180 - 320\grad, the
same place where the helium abundance is the weakest. These abundance
variations (with rotational phase) support predictions made by
the theory of atomic diffusion in the presence of a magnetic field.
Simultaneously, these result support the possibility of the formation of unusual
structures in stellar magnetic fields. Analysis of vertical
stratification of the silicon and helium abundances shows that the
boundaries of an abundance jump (in the two step model) are similar
for each element; 
$\tau_{5000}$ = 0.8-1.2 for helium and 0.5-1.3 for silicon.
The elemental abundances in the layers of effective formation of
selected absorption lines for various phases are also correlated
with the excitation energies of low transition levels: abundances are
enhanced for higher excitation energy and higher optical depth
within the applied model atmosphere.

\end{abstract}


\section{Introduction}
HD~21699 (HR~1063) was initially classified by Roman and Morgan
(1950) as a B8IIIvar star. Molnar (1972) later performed an analysis
of its spectra and determined that helium was extremely deficient
(by factor of 5), as a result he classified this star as a He-weak.
Shore et al. (1987) refer to HD~21699 as a He-weak silicon star (sn
class - with broad and diffuse He I lines). The Sn class is
primarily attributed by Morgan (1977) to the silicon and He-weak
star HD 5737. In the classification of Preston (1974) both objects
are categorised as chemically peculiar (CP2) stars. Because of the
extremely reduced helium abundance, it is possible that HD~21699 has
been given an incorrect spectral classification: the MK
classification gives Sp = B8, whereas Abt et al. (2002) suggest a
spectral class of B8IIIpMn that corresponds to \Tef = 12000 K, while
Glagolevskij (2002) derived \Tef = 16100 K from the analysis of
colour indices. Using the spectra obtained with the 6-m telescope,
the following parameters were derived by Glagolevskij et al. (2006) by
an analysis of the $H_{\delta}$ line: \Tef = 16000 K, lg g = 4.15,
\Vt = 0.8 km\,$s^{-1}$.

The period of axial rotation for HD~21699 is P = $2^d.4765$ (Brown et al. 1985).
Using the relation between the equatorial velocity (km\,$s^{-1}$), period (days),
stellar radius in solar radii ($V =  50.613* R/P$) and the measured value of \vsini =  35 km\,$s^{-1}$
yields an inclination angle of $i$ = $32^{\circ}$ 
(Glagolevskij \& Chuntonov 2007). The positive magnetic field maximum occurs at the phase $\phi= 0$, and the negative
extremum occurs at $\phi=  0.4$, according to the ephemerides of
Glagolevskij \& Chuntonov (2007) for initial phase; (JD = 2445595.529 + $2^d.49246$)

\vskip2mm

\section{Magnetic field model and variability of He I and Si II lines}

\vskip2mm

It has been shown by Stateva (1995) that a one-spot model for the
helium and silicon surface distributions fits the observed periodic
equivalent width variations in both He I lines and Si II lines. In
this case one large He-weak spot is situated around the positive
magnetic pole, while the Si II spot is located at the negative pole,
with the surface magnetic field assumed to be due to a centered
dipolar field. Analysis of the equivalent widths of the helium and
silicon lines show that the maximum values do not occur at the same
phase and are in fact opposite. Usually, in a case of centered
magnetic dipole the abundances of the same element are equal in the
vicinity of both magnetic poles. UBV photometry (Percy 1985) shows
only a single ''peak'' in the light curve during
one complete rotational period; typically stars with a centered
dipolar magnetic field display a double wave (one peak and trough).
In the work of Shore et al. (1987) it is mentioned that a similar
behavior is noted in the photometric data of two other He-weak sn
stars: HD~5737 and HD~79158.

Brown et al. (1985) reported the discovery of magnetically
controlled stellar mass outflow in HD~21699 based on IUE
observations of the C IV resonance doublet, which is variable on the
rotational time-scale of about 2.5 days, but found only one jet
from one of the magnetic poles.

An additional study of the surface magnetic field for this star was
performed by Glagolevskij \& Chuntonov (2007) using a model of the
spatially distributed magnetic charges (Gerth \& Glagolevskij,
2001). This approach differs from other methods and could be
considered more representative of the real physics because the magnetic
field should have a vortex current source. Varying the necessary
parameters, authors have calculated phase curves of the mean
effective field $B_e$ and of the mean surface magnetic field $B_s$,
which have then been compared with observed data. By the method of
successive approximation (step-by-step approach) they have found the
best fit between the calculated data and the observations.

Measurements of the $H_{\beta}$ line with the Zeeman polarimeter
(Brown et al. 1985) were employed to constrain the magnetic field
model. A better agreement between the observed and calculated
equivalent widths curves of He I and Si II lines for HD~21699 is
reached with a model where the dipole is displaced across its
magnetic axis from the stellar center by a distance $a$ = 0.4 $\pm$
0.1, expressed in stellar radii units. Therefore in this
configuration the magnetic poles appear to be close to one another
on the stellar surface (they are separated by $55^{\circ}$, not by
$180^{\circ}$, as it would be in the case of a centered dipole). The
big "strong" helium spot is situated at the average position of the
two magnetic poles, while the "weak" spot is located on the opposite
side of the star (Glagolevskij \& Chuntonov 2007).


The equivalent widths of He I and Si II lines show extreme values at
phases that correspond to the passage through the visible meridian of
zero magnetic field between the magnetic poles. The intensity of the
helium line reaches its maximum at the magnetic poles, while the
intensity of the silicon lines reaches its minimum in the same
place. Silicon abundance is maximal in the regions where the magnetic
field is predominantly tangential to the stellar surface.

The behavior of atomic diffusion does not depend on the sign of the
magnetic field nor does the abundance of chemical species in a
magnetic spot depend on the field's sign. Observational data shows
that despite the helium lines being weakened in all rotational
phases, the helium is clearly enhanced at the magnetic spot and on
the contrary, silicon lines are weaker in that spot than on the
opposite part of the stellar surface. Due to averaging over the visible
hemisphere and owing to the close location of both magnetic poles to
each other, the only intensity variation (one wave) in spectral
lines is observed for the two investigated chemical species. For the
same reason we observe only one wave variability in the
photometric light curve and for average surface magnetic field
$B_s$. Respectively, only one stellar wind jet, common from both
poles, is observed.


HD~21699 is a unique star when taking into account the nature of
the surface magnetic field distribution. Due to the close proximetry of
the positive and negative magnetic poles to each other, the common
"magnetic spot" is situated on the half of the stellar surface with
predominantly vertical magnetic field lines (Glagolevskij \&
Chuntonov, 2007), while in the other hemisphere magnetic field
lines are horizontal. Such a magnetic field configuration provides
an excellent ''lab'' for studying abundance surface distributions
for chemical species concentrated around magnetic poles (He I and
others) and those that are concentrated in areas with horizontal
magnetic field lines (Si II and others). It also provides a unique
opportunity to study the variability of mean abundances with
rotational phase, and the vertical distribution (stratification) of
chemical species in stellar atmospheres depending on the structure of the
magnetic field.

HD~21699 is a very suitable star for studying the atomic diffusion
of chemical species in the atmospheres of CP stars because its
surface can be simply divided in two parts: one, where the magnetic
field lines are horizontal to the stellar surface, and another,
where they are predominantly vertical.

\section{Observation data and treatment}

Medium-resolution spectra were obtained using the main stellar
spectrograph (MSS) (equipped with a slicer of 14 cuts and spectral
resolution R = 15000) on the 6-m telescope of the Special
Astrophysical Observatory of Russian Academy of Sciences (SAO RAS).
Nine spectra in the region 3900-4300 \AA\ were obtained to uniformly cover the whole rotational period, with a signal-to-noise ratio
of 2000. One additional higher resolution spectrum was obtained with
the Nasmyth Echelle Spectrometer (NES, see Panchuk et al. 2002) for a 
resolution of 40000 in
the range 4500-5900 \AA\, (the phase=0.687). The signal-to-noise
ratio for this spectrum is 300. The SAO spectra are reduced using
MIDAS procedures.

We also used spectropolarimetric data (Stokes I and V) obtained with
the MuSiCoS (MUlti SIte COntinuous Spectroscopy) spectropolarimeter which was installed on the 2-m Bernard Lyot telescope at
the Pic du Midi Observatory in France. MuSiCoS is a table-top
cross-dispersed echelle spectrograph which is fed by two optical
fibres from the polarimeter, mounted in the cassegrain focus, with
resolution R = 35000 and wavelength coverage from 4500 to 6600 \AA\,
(phases=0.118, 0.322, 0.525). The signal-to-noise ratio for the
MuSiCoS spectra is about 400 per pixel.

A complete Stokes exposure is made up of 4 subexposures in which the
retarder is rotated by 90 degrees and back. In principle any first order spurious
polarization signatures are suppressed to an acceptable level, by switching the beams
within the instrument. The MuSiCoS spectra are reduced using the ESPRIT data reduction package
(Donati et al. 1997). We only used Stokes I spectra from the MuSiCoS
data, which were obtained with observing procedures described in
detail by Donati et al. (1999). Stokes V spectra are not discussed
here simply because the three available phases are not
sufficient to provide additional information about the magnetic
field geometry.

\begin{table}

 \centering


  \caption{\label{_tab1}  Data of observed spectra (JD, wavelength coverage, S/N and
rotational phase)}

  \begin{tabular}{ccccc}


\hline

\hline

                     &                      &                    &        &       \\

   JD                &         spectrograph &    wavelength cov. &    S/N & phase \\

                     &                      &                    &        &        \\

\hline

\hline


2453591.6342    &            MuSiCos      &   4490 - 6620 \AA  &   300-410  &   0.118 \\

2453594.6357    &             "        &     "              &  340-450   &  0.322  \\

2453607.6030    &             "         &     "             &  240-360   &  0.525 \\

2453605.510     &              NES       &    4500 - 5900 \AA    &    300     &   0.687 \\

2454072.444     &              MSS        &   4000 - 4240 \AA     &   2000     & 0.023 \\

2454070.182     &              "         &      "            &    "     &   0.116 \\

2454464.188     &              "           &    "                 &   "        &   0.195 \\

2454073.167     &              "           &    "                 &   "        &   0.314 \\

2454462.189     &              "           &    "                 &   "        &   0.393 \\

2454073.558     &              "           &    "                 &   "        &   0.470 \\

2454101.155     &              "           &    "                 &   "       &    0.543 \\

2454485.251     &              "           &    "                 &   "        &   0.646 \\

2454072.208     &              "           &    "                 &   "         &  0.928 \\

                &                         &                      &            &        \\

\noalign{\smallskip}

\hline

\end{tabular}


\end{table}


In Table 1 we provide the JD, wavelength coverage, S/N and
rotational phase for all spectra. The nine spectra obtained with MSS SAO
RAS are shown in Fig.1, where we can see variation in intensity
of He I 4026 \AA, Si II 4128~\AA\,and 4130 \AA\, lines during
rotational period.

\begin{figure}

\includegraphics [width=85mm]{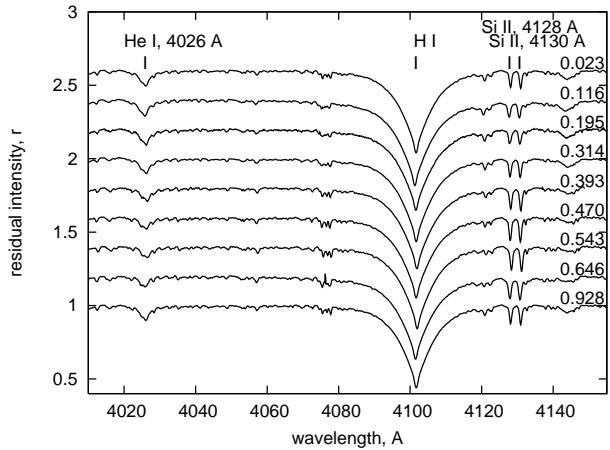}

\caption[]{Nine spectra of HD 21699 obtained with MSS SAO RAS. We
can see variation of intensity of He I 4026~\AA, Si II 4128~\AA\,
and 4130~\AA\, lines during rotational period.}

\end{figure}

\section{The spectra modeling}

Synthetic spectra for HD~21699 were calculated with the code SYNTHM
of Khan (2004) and SYNTHV of Tsymbal (1996), both codes calculate
the line profiles or synthetic spectra formed from the entire visible
hemisphere of the star. We used Kurucz's model atmospheres (1994) as
well as Pavlenko's (2003) model atmospheres with a reduced He
abundance. The atomic lists of VALD (Kupka et al.1999) and Castelli
(http://wwwuser.oat.ts.astro.it/castelli/) were used as the input
atomic data for the spectrum synthesis. The average abundance of
each chemical species (assuming no vertical stratification) was
determined by comparing observed spectral profiles to those produced
by the models: SYNTHV code (no magnetic splitting) and SYNTHM code, that 
takes into account magnetic splitting of lines (Version-04 without 
stratification).
Stratification of chemical elements was determined in a similar vein
using the SYNTHM code (Version-05). Previously, we have determined
the contribution function for each line using the code WITA
(Pavlenko 1997).

\section{RESULTS}

{\bf HELIUM}: The helium lines are weakened in the spectra of
HD~21699. Osmer and Peterson (1974) and Vauclair (1975) have shown
that the formation of He-weak or He-rich stars depends on the wind power
in their atmospheres. If we define VW as the relative velocity of
flux with respect to a stellar wind and VD as the velocity of
diffusion inside a star, then He-weak stars are formed when VD $>$
VW (Vauclair 1975). It is clear from Fig. 2 that over the whole
surface of HD 21699, helium appears to be deficient. Nevertheless,
in the area of the "magnetic spot" the helium abundance is
comparatively higher (by a factor 1.5), than what is derived for the
opposite side.

In Fig. 2 we show the variation of helium abundance with the
rotational phase. Circles represent the averaged abundance obtained
from analysis of four He~I  lines (4026 \AA\ in SAO MSS spectra and
4713, 4921, 5015 \AA\ in the MuSiCoS spectra), while the squares
represent an abundance deduced only from one He I line (4026 \AA\ )
using SAO MSS spectra). Errors in the averaged He abundances,
estimated from the standard deviation of the data obtained for all
analyzed He lines at respective rotational phases, are specified by
vertical bars.
The uncertainty for the
single He I 4026 \AA\ abundances for all phases was estimated as
0.1 dex. All the results for Fig.~2 were obtained using SYNTHM code
which takes into account magnetic splitting of spectral lines. The
abundance estimates are derived for the visible hemisphere for each
phase. 

\begin{figure}

\includegraphics [width=65mm, angle=0]{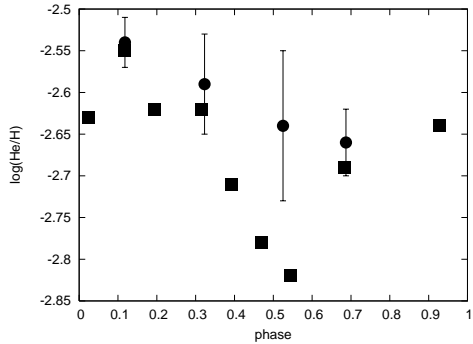}

\caption[]{Variability of the helium abundance (log N(He)/N(H)) with
rotational phase. The circles indicate abundances deduced from
analysis of four He I lines (MuSiCoS and NES spectra), while the
squares indicate abundances deduced from analysis of the He I line
4026 \AA\,. All abundance values are averaged over the visible
hemisphere for each phase (see Table.1), not taking into account
helium stratification in the atmosphere and presence of spots.
}

\end{figure}

For those parts of the stellar surface where the magnetic lines are
horizontal, only vertical diffusion of He I (not He II) is
efficient. The helium abundance in the aforementioned parts of
stellar surface must be higher than at the magnetic poles, where the
field lines are vertical, because there is no resistance to
diffusion by both the He I and He II atoms. The fact that the
intensity of He I lines at the magnetic poles is higher than on
the opposite side supports the idea of a sufficiently strong wind
at the magnetic poles. The observed data justify the presence of a
powerful wind from the "magnetic spot" of HD~21699 (magnetically
structured jets, Brown et al. 1985). We have assumed the two-step
approximation for vertical stratification of chemical species in the
atmosphere of HD~21699 (see, for example, Ryabchikova et al. 2005).

For the lower layers, the abundance was derived from the line wings,
while for the upper layers it was derived from the line cores. { An
example of the fit for the He I 4026 \AA\ line profile at phase
0.116 (MSS spectrum) is shown in Fig.~3. It is obvious that the line
wings are quite sensitive to the helium abundances at the lower
atmospheric layers. Therefore, the respective helium abundance can
be specified with high confidence, namely with an error of 0.1 dex.
Meanwhile, the core of a line provides information about the element
abundance in the upper atmospheric layers. However, it provides a
somewhat larger error, 0.2 dex. Only by taking into account the He
stratification we can get a good fit between the model and the
observed diffuse He line profiles for this He-weak sn-type star.

In Fig.~4 we show the vertical stratification of helium for nine phases,
derived from the analysis of the He I 4026 \AA\, line profiles using
the Kurucz model atmosphere 16000/4.0 for the visible hemisphere
for each phase.

\begin{figure}
\includegraphics [width=65mm, angle=0]{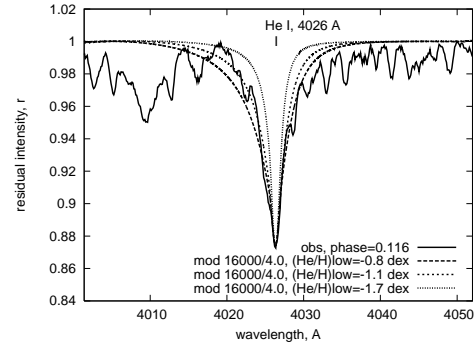}
\caption[]{The fits of calculated He I 4026 \AA\ line profiles to observed one for phase 0.116.
Here (He/H)low specifies helium abundance at the lower atmospheric layers assuming the two steps
model for helium vertical stratification.
}
\end{figure}

\begin{figure}
\includegraphics [width=65mm, angle=0]{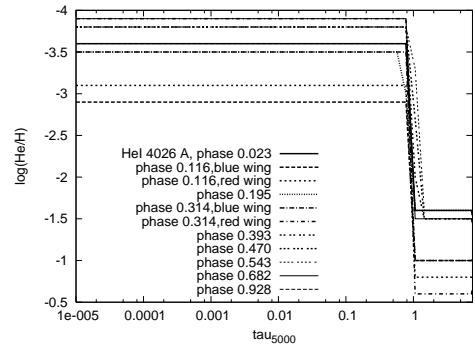}
\caption{Helium vertical stratification (in the frame of two-step
model) from the analysis of He I 4026 \AA\ line for various
rotational phases. In phases 0.928, 0.023, 0.116, 0.195, 0.314
helium abundance is close to its maximum, while in phases 0.393,
0.470, 0.543, 0.682 helium abundance is close to its minimum.}
\end{figure}

\begin{figure}
\includegraphics [width=65mm, angle=0]{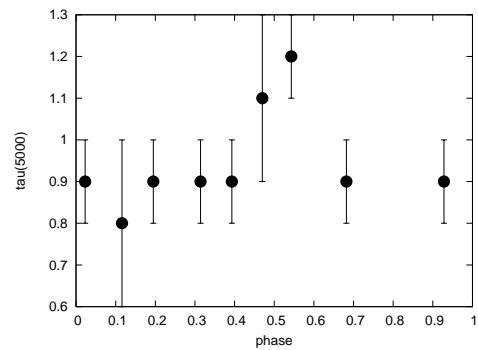}
\caption[]{Optical depth $\tau_{5000}$ of the abundance jump with bars of errors versus rotational phase
for the line He I 4026 \AA\ .}
\end{figure}

The results shown in Fig.~4 confirm the theoretical prediction of
Vauclair et al. (1991), which predict the enhancement of helium
abundance towards a deeper optical depth. In Fig.~5, we can see that
as the phase changes, the optical depth $\tau_{5000}$ correspondes
to where the abundance jump behaves inversely to the variation of the
He abundance (see Fig.2).


{\bf SILICON}: In Table 2, we show estimates of the silicon
abundance derived from MuSiCoS and NES spectra for seven Si II lines
with various excitation energies (in the range of 8-16 eV) of the
lower transition level for 4 rotation phases. These values were
obtained for the visible hemisphere for each phase without taking
into account the silicon stratification in the atmosphere of the
star. It is easy to track an apparent increase of the derived
silicon abundance with the rise of excitation energy (i.e., towards
the deeper atmosphere). This tendency supports the idea of vertical
stratification of silicon as suggested by Vauclair et al. (1979).

\begin{table}

\caption{Variation of Si II mean abundance for 4 rotational phases: 0.118, 
0.322, 0.525 are from MuSiCoS spectra and 0.687 is from NES spectrum. 
}

\begin{tabular}{|c|c|c|c|c|c|c|}

\noalign{\smallskip}

\noalign{\smallskip}

\hline

\noalign{\smallskip}

\noalign{\smallskip}

line (A)& EP(eV)& \multicolumn{4}{c}{log N(SiII)/N(H)} vs. phase \\

        &   &     0.118 &  0.322 &    0.525 &    0.687  \\

\noalign{\smallskip}

\noalign{\smallskip}

\hline

\noalign{\smallskip}

\noalign{\smallskip}

6347 &    8.12  &     -5.12  &  -4.92  &   -4.02  &    -     \\

6371 &    8.12  &     -5.12  &  -4.92  &   -4.12  &    -     \\

5041 &   10.07  &     -4.87  &  -4.67  &   -4.02  &   -4.25  \\

5055 &   10.07  &     -5.15  &  -4.99  &   -4.55  &   -4.50  \\

4673 &   12.84  &     -3.87  &  -3.82  &   -3.52  &   -3.70  \\

5669 &   14.21  &     -4.15  &  -4.05  &   -3.55  &   -3.75  \\

5202 &   16.35  &     -3.75  &  -3.80  &   -3.20  &   -3.70  \\

\noalign{\smallskip}

\noalign{\smallskip}

\hline
\end{tabular}
\end{table}

\begin{table}

\caption{Variation of mean silicon abundance log N(SiII)/N(H)
(without stratification) with rotational phase (including MSS
spectra): the phases 0.118, 0.322 and 0.525 A are taken from MuSiCoS spectra, the phase 
0.687 is NES spectrum. Wavelengths and EP of spectral lines are given in 
Table 2.}

\begin{tabular}{|c|c|c|c|c|c|}

\noalign{\smallskip}

\noalign{\smallskip}

\hline

\noalign{\smallskip}

\noalign{\smallskip}


Phase &  mean 4128,  & \multicolumn{3}{c}{mean (MuSiCoS \& NES) } \\

      &  4130\AA, EP=9.84 &   8-10eV  & 12-16eV & Si III 4552\AA  \\


\noalign{\smallskip}

\noalign{\smallskip}

\hline

\noalign{\smallskip}

\noalign{\smallskip}

0.023    &          -4.90  &                             &            & \\

0.116    &          -4.95  &                             &            & \\

0.118    &            -    &                     -5.06   &      -3.92 & -3.77 \\

0.195    &          -5.15  &                             &            & \\

0.314    &          -4.80  &                             &            & \\

0.322    &            -    &                     -4.88   &      -3.89 & -3.75 \\

0.393    &          -4.50  &                             &            & \\

0.470    &          -4.35  &                             &            & \\

0.525    &             -   &                      -4.16  &       -3.42& -3.77 \\

0.543    &          -4.35  &                             &            & \\

0.682    &          -4.15  &                             &            & \\

0.687    &             -   &                      -4.37  &       -3.72& \\

0.928    &          -4.70  &                             &            & \\

\noalign{\smallskip}

\noalign{\smallskip}

\hline

\end{tabular}

\end{table}

\begin{figure}

\includegraphics [width=65mm, angle=0]{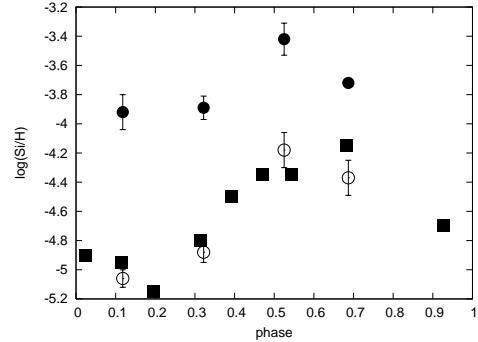}

\caption[]{ Variation of silicon (Si II) abundance with rotational
phase. The estimates of log N(Si)/N(H) are derived from the lines of
low excitation energies (clear circles stand for eshelle, MuSiCoS
and NES spectra, while shaded squares stand  for MSS spectra) and
high excitation energies(shaded circles stand for eshelle spectra).
Vertical bars show errors of the averaged estimates. The error for
single abundances was estimated as 0.1 dex}

\end{figure}

Table~3 shows a combination of the aforementioned data with the
silicon abundances derived from Si II lines 4128 \AA\, and 4130
\AA\, for nine rotational phases using the MSS spectra. Fig.~6
presents them graphically as a variation of silicon (Si II)
abundance with rotational phase.  These abundance estimates, like He
I abundances for all phases, were obtained with the SYNTHM code taking
into account magnetic splitting for values $B_s$ corresponding to
each phase from the magnetic field model. We see that silicon has a
higher abundance where the surface magnetic field has the lowest
value (phases 0.525, 0.543, 0.682).

Taking into account the level (-4.49 dex) of solar abundance for
silicon (Grevesse et al. 2007), it appears that the lines with  low
excitation energies show an abundance deficit, while the lines with
high excitation energy show an excess of silicon for {the phases
with maximal $B_s$ (0.118 and 0.322).

 We also determined Si III abundance from the 4552 \AA\, line using
the same model atmosphere 16000/4.0 (see last column of Table~3) for
the three phases of MuSiCoS spectra. The results are close to abundances
derived from the Si II lines of high excitation energies, justifying the
choice of the model atmosphere.

It is remarkable that the abundance estimates derived from the
analysis of lines with low excitation energies are in good agreement
among themselves for all available spectra (MuSiCoS, NES and MSS).
For the phases with minimal $B_s$ (0.525 and 0.687) all the lines
show an excess of silicon. Nevertheless, the lines with higher
excitation energies still result in higher silicon abundance. They
formed at deeper atmospheric layers and as a result we observe an
enhancement of silicon abundance there. This is most obvious in the
case of the "magnetic spot". In Fig.7 we show the fit of Si II
lines 4128 and 4130 \AA\ to model profiles, calculated by taking into
account magnetic broadening and stratification of silicon in  the
atmosphere. Note the relatively good fits of profiles.

\begin{figure}

\includegraphics [width=65mm, angle=0]{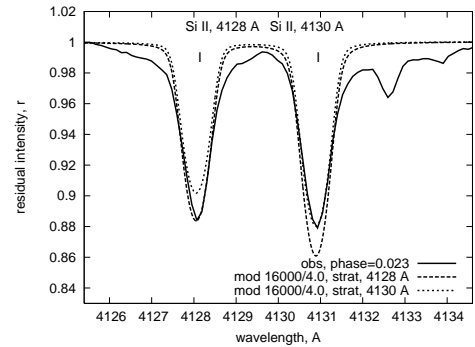}

\caption[]{The fit of Si II lines 4128 and 4130 \AA\ to model
profiles, calculated taking into account magnetic broadening and
stratification of silicon in the atmosphere.}

\end{figure}

\begin{figure}

\includegraphics [width=65mm, angle=270]{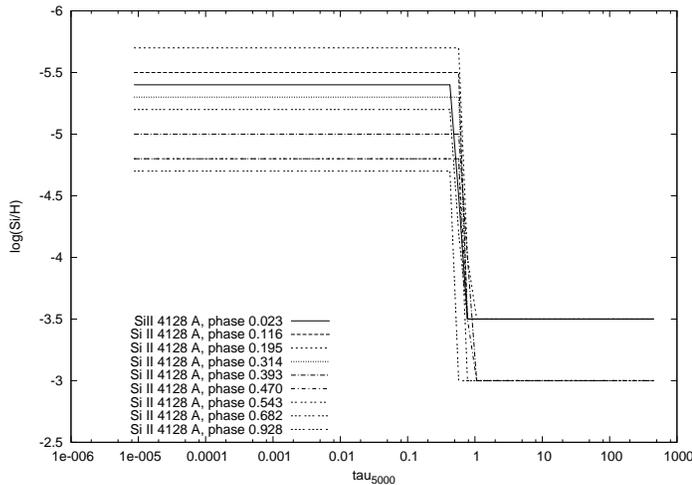}

\caption[]{Vertical stratification of silicon deduced from analysis
of Si II line 4128 \AA\  assuming the two-steps model. Phases 0.023,
0.116, 0.195, 0.314,0.928 correspond to minimal Si II abundance
(maximal He I), while phases 0.393, 0.410, 0.470, 0.543, 0.682
correspond to maximal Si II abundance (minimal He I).}

\end{figure}

Fig.~8  shows an enhancement of silicon abundance towards the deeper
atmospheric layers, as predicted by Vauclair et al. (1979). Fig. 9
presents stratification of Si II derived from lines with different
excitation energies for the phase of minimal magnetic field.

\begin{figure}
\includegraphics [width=65mm, angle=270]{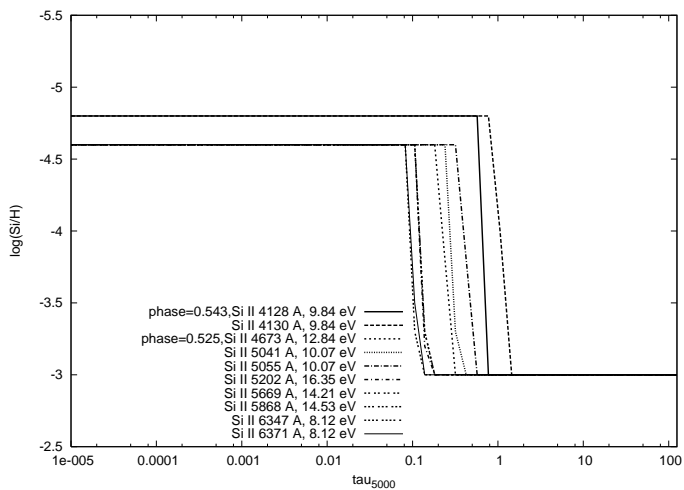}
\caption[]{Vertical stratification of silicon for the phase 0.525, where $B_s$ reaches its minimum
and Si II abundance rises to its maximum.}
\end{figure}

The case of non-uniform silicon distribution on the surface of
magnetic stars is discussed in the works of Vauclair et al. (1979),
Alecian and Vauclair (1981) and Megessier (1984). Silicon usually
accumulates in the places where the magnetic field lines are
predominantly horizontal and where they can oppose the gravitational
settling of ionized silicon. In the case of a shifted magnetic
dipole, like HD 21699, the side opposite to the magnetic poles has a
large area with horizontal magnetic field lines, where silicon
should be concentrated. Meanwhile, it has to be weakened around the
magnetic poles.

From Tables~2 and 3 and Fig.~6, 7 and 8 it appears that our results
confirm the predictions of Vauclair et al. (1979), Alecian and
Vauclair (1981) and Megessier (1984). In the atmosphere of HD 21699
silicon is enhanced in the area where the field lines are horizontal
to the stellar surface. The optical depth of the abundance jump for
silicon $\tau_{5000}$ (like to helium) has a tendency to change with
phases reversely to the silicon abundance variation (see Figs. 6, 10).


\begin{figure}

\includegraphics [width=65mm, angle=0]{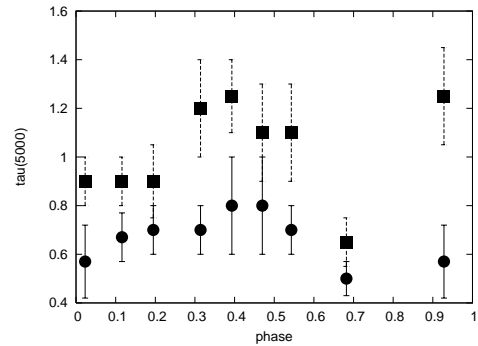}

\caption[]{Optical depth $\tau_{5000}$ of the abundance jump versus
rotational phase for the Si II lines 4128 \AA\ (filled squares) and
4130 \AA\ (filled circles) with bars of error. As in the case of He
I, we can see a tendency to change $\tau_{5000}$ with phase contrary
to abundances.}
\end{figure}

Abundance stratification due to diffusion processes acting in the atmospheres
of chemically peculiar stars is studied in the recent work of Monin \& LeBlanc (2007).
Their self-consistant models show that elements such as Fe, Cr, Si and Ca do indeed 
accumulate at large optical depths, while they are dramatically underabundant in the upper atmosphere.
The transition zone for iron in their models is located around optical depth $\tau_{5000}$ = 1.

\section{Basic conclusions}
For the first time photospheric chemical element abundances of HD
21699 were obtained during the whole rotational period. { Magnetic splitting of spectral lines and elements
stratification in the atmosphere were taken into account with line
profile modelling. Helium abundance is weakened for the whole
stellar surface, which is usual for He-weak stars. Nevertheless,
helium has higher abundance in the region of the "magnetic spot" due
to the influence of stellar wind, which elevates the helium to the
outer layers of the stellar atmosphere, and helium abundance should
increase with optical depth in the atmospheres of He-weak stars.
(Vauclair et al. 1991).

Silicon is accumulated in the part of the star where the magnetic
field lines are predominantly horizontal, as it was predicted by
Vauclair et al. (1979), Alecian and Vauclair (1981)  and Meggessier
(1984). Silicon abundance determined from lines with low
excitation energies (8-10eV) appears to be lower in the area of the
magnetic poles and is approximately solar in the region with
horizontal magnetic lines. The lines with high excitation energies
(12-16eV) also show enhancement of silicon abundance for the region
with horizontal magnetic lines, but this abundance is significantly
higher than solar everywhere on the stellar surface. Silicon
abundance is lower in the outer parts of the atmosphere and higher
in the deeper layers.

The optical depth of the abundance jump (in two-steps model) is near
to  $\tau_{5000}$ = 1. In the cases of He I and Si II, we can see
a tendency to change $\tau_{5000}$ with the phase contrary to
abundances, which could be the result of competition between the stellar
wind, diffusion, and the local magnetic field geometry that would lead to
accumulations of helium and silicon at different depths in the
atmosphere.

Since IUE observations of the C IV resonance doublet show existence of
magnetically controlled stellar wind in HD 21699 (Brown et al. 1985), one
should expect some variations in Balmer lines, similar to that seen
in H-alpha in 36 Lyn (Wade et al. 2006) and in HD 37479 (Smith et al.
2006) and in many Balmer lines in HD 37479 (Smith and Bohlender, 2007).
Unfortunately, due to the incomplete number of observed phases it is
difficult to find similar changes in Balmer lines in HD 21699.

\section{aknowledgements}
We thank V. Tsymbal and S. Khan for the use of their codes: SYNTHV
and SYNTHM. Many thanks to the referee of the paper, Dr. Bohlender for
the very valuable advice and comments. This work was partially
funded by the Microcosmophysics program of National Academy of
Sciences and National Space Agency of Ukraine and by the Natural
Sciences and Engineering Research Council of Canada (NSERC).

\end{document}